\documentclass[useAMS,usenatbib]{mn2e}
\usepackage{graphicx}

\title[Dust properties in Taurus-Auriga]
  {Variation of the  ultraviolet extinction law across the Taurus-Auriga star forming complex.\\
A GALEX based study.}
\author[A.I. G\'omez de Castro et al.]
  {Ana I. G\'omez de Castro$^1$,
  Javier L\'opez-Santiago$^1$, F\'atima L\'opez-Mart\'{i}nez$^1$,  
\newauthor  N\'estor S\'anchez$^1$, Elisa de Castro$^2$ and  Manuel Cornide$^2$ \\
  $^1$AEGORA Research Group, Universidad Complutense de Madrid, Plaza de Ciencias 3,
28040 Madrid, Spain\\
  $^2$Fac. de CC. F\'{\i}sicas, Universidad Complutense de Madrid, Plaza de Ciencias 1,
28040 Madrid, Spain}
\date{Submission, October 15th, 2013}

\pagerange{\pageref{firstpage}--\pageref{lastpage}} \pubyear{}

\begin{document}

\maketitle

\label{firstpage}

\begin{abstract}

The Taurus-Auriga molecular complex (TMC) is the main laboratory for the study of low mass star formation. 
The density and properties of interstellar dust are expected to vary across the TMC. These variations trace
important processes such as dust nucleation or the magnetic field coupling with the cloud. 
In this article, we show how the combination of near ultraviolet (NUV) and infrared (IR) photometry can be used
to derive the strength of the 2175~\AA\ bump and thus any enhancement in the abundance of small dust grains and PAHs
in the dust grains size distribution. This technique is applied to the envelope of the TMC, mapped by the GALEX 
All Sky Survey (AIS).   UV and IR {\bf photometric} data have been retrieved from the GALEX-AIS and the 2MASS catalogues. 
NUV and K-band star counts have been used to identify the areas in the cloud envelope where the 2175~\AA\ bump is weaker
than in the diffuse ISM namely, the low column density extensions of L1495, L1498 and L1524 in Taurus,  L1545, L1548, L1519,
L1513 in Auriga and L1482-83 in the California region. This finding agrees with previous results on dust evolution
derived from Spitzer data and suggests that dust grains begin to decouple from the environmental galactic magnetic field 
already in the envelope.

\end{abstract}


\section{Introduction}

Molecular clouds are not isolated entities in the galactic interstellar medium. They are produced at locations where diffuse gas in the Galaxy is gathered due to the action of galactic stresses basically, spiral waves exciting Parker-Jeans instability modes (Elmegreen 1990, Franco et al. 2002) though eventually accretion from the intergalactic medium cannot be neglected (Mirabel 1982, Wang et al. 2004).

Molecular clouds form in large structures or {\it molecular complexes} in specific areas of the Galaxy. The Taurus-Auriga Molecular Complex (TMC) covers about 20$ ^{\rm o}$x20$ ^{\rm o}$ in the sky.
The TMC  is the prototype of filamentary molecular complex where star formation proceeds less efficiently than in massive clusters, such as the L1630 cloud in the Orion molecular complex (Lada 19922). The large filamentary structure of the TMC is reminiscent of the galactic open clusters distribution.

Molecular clouds are submitted to the ionising radiation field of the Galaxy that produces photo evaporative flows on the surface. The hardness of the ultraviolet (UV) radiation field around the TMC can be readily inferred from the SPEAR/FIMS imaging spectrograph observations. The cloud core and the halo region are clearly identified {\bf (Lee et al. 2006)}. The scattered far UV(1370\AA\ -- 1670\AA ) radiation is seen in, and beyond, the halo region (where $A_v <1.5$~mag) as plumes in photo-evaporative flows, on the scale of the SPEAR/FIMS resolution: 5-10~arcmin or 0.2-0.4~pc . Mapping the diffuse emission and the extinction variations at those optical depths provides important information on the physics of the cloud boundary that, contrary to the cloud interior, is not dominated by the gravitational field but by more subtle stresses. Moreover, there is recent evidence of additional heating sources in the edge of the TMC, since the diffuse galactic UV radiation field cannot solely account for the degree of excitation of the molecular Hydrogen (Goldsmith et al. 2010). 

Cloud haloes may provide important clues on the transmission/reflection of long wavelength modes of the galactic magnetic field, as those hypothesised by G\'omez de Castro \& Pudritz (1992, hereafter GdCP92) to describe the global properties of the Taurus star-forming region. The geometry of the magnetic field in Taurus has been traced by optical and infrared polarisation measurements (Vrba et al. 1976, Moneti et al. 1984, Heyer et al. 1987, Goodman et al. 1990, Chapman et al. 2011). The observations indicate that the field is well ordered on the scale of the whole cloud, but that it also contains a significant disordered component on the smaller scales (Goodman et al. 1990).  Zeeman measurements of the field strength in molecular clouds indicate that magnetic and gravitational energy density are comparable. Thus, magnetic fields must play a fundamental role in the dynamics of the cloud and the formation of structures such as the ubiquitous cores and filaments. The orientation of the magnetic field with respect to the dense filament substructure can vary from perpendicular to parallel alignments depending upon the filament (Vrba et al. 1976, Heyer et al. 1987, Goodman et al. 1990). A recent evaluation of the field relevance in the gas flow suggests that magnetic fields control the mass flow in the TMC diffuse envelope. Heyer \& Brunt (2012) studied the alignment of the velocity anisotropy with the magnetic field over varying physical conditions and environments to conclude that the velocity anisotropy is aligned with the local, projected mean magnetic field direction in low surface brightness $^{12}$CO emission areas, corresponding to regions of low visual extinction and presumably, low gas volume density. This would be consistent with the presence of a large scale Alfv\'en wave channelling the dynamics of the TMC, as suggested by GdCP92. 

The coupling between the clouds and the ambient field relies on the dust grains ionisation and mass spectrum. Dust grains contain most of the mass of the charged particle component in the molecular clouds interior. For typical clouds, the cut-off wavelength for the coupling with the ambient field decreases from 1~pc to 0.1~pc when dust grains are considered (Nakano, 1998). Also, the size of the dust grains is relevant for the resonance frequencies to Alfv\'en waves propagation and the neutral-charge particles coupling (Pilipp et al. 1987).  The near ultraviolet bump (the 2175 \AA\ bump) is the most sensitive feature to the presence of small dust grains (see i.e. Draine 2003) and hence, to plasma field coupling.  Therefore, though the use of spectral tracers able to penetrate the dusty environment (infrared, X-ray, radio-wavelengths) is useful to study the gravitational collapse and the cloud fragmentation, important pieces of information leading to cloud formation and support are best accessed through UV observations. 

The GALactic Evolution eXplorer (GALEX) has surveyed the envelope of the TMC complex permitting {\bf a study of } the dust spatial distribution with a sensitivity significantly higher than the conventional means (molecular or infrared mapping). Moreover, as the TMC is a well studied region, comprehensive mapping in various molecular probes such as CO (Ungerechts \& Thaddeus 1987, Heyer et al. 1987, Goldsmith et al. 2008, Davis et al. 2010), NH$_3$ (Benson \& Myers, 1983, Gaida et al. 1984, Olano et al. 1988, Ladd et al. 1994) as well as in the infrared range (Froebrich et al. 2007, hereafter F07, Lombardi et al. 2010, hereafter LLA10) is available allowing to study the relation between gas and dust density in various bands. For instance, the comparison between the relative extinctions in the GALEX NUV band, centred at the 2175 \AA\ bump, and in the K band from the 2MASS survey (LLA10) permits to evaluate variations in the average dust grains size over the TMC in a very efficient manner.  Evidence of dust grain growth in dense molecular gas filaments compared with the diffuse interstellar medium (ISM) have been recently, reported from infrared studies  (see {\it e.g.} Ysard et al 2013 study for the L1506 filament in Taurus or Flagey et al. 2009 for a Spitzer based study).

The UV bump is, by far, the strongest spectral feature in the extinction curve but its source remains uncertain (see {\it e.g.} the review by Draine 2003{\bf )}. Though small graphite grains  were
proposed initially as the main source of the bump, the baseline today are polycyclic aromatic hydrocarbon (PAH) molecules (Weintgarner \& Draine, 2001), that share with a graphite sheet, a similar structure in terms of the distribution of the  carbon atoms. PAHs are required to reproduce the observed infrared emission (Leger \& Puget, 1984)  and they somewhat represent the extension of the dust grain size distribution into the molecular domain. 

In this article, we derive  a method to determine the strength of the 2175~\AA\ bump from a combination of extinction measurements in the GALEX near UV (NUV) band and the 2MASS K infrared band (see Sect.~2). In Section~3, we describe the GALEX All Sky Survey  of the TMC, the quality and main characteristics of the data. In Section~4, we describe the method used to measure the
extinction in the NUV band in the TMC and the results are analysed in Section~5. 
It is shown that the strength 2175~\AA\ bump decreases with respect to the average ISM values in
the filaments gathering areas, namely, the low column density extensions of L1495, L1498 and L1524 in Taurus,  L1545, L1548, L1519,L1513 in Auriga and L1482-83 in the California region. These finding agrees with previous results on dust evolution
derived from Spitzer data. This indicates that within the halo of the molecular clouds, the dust grains are already rather large, affecting the coupling between the dust and the environmental magnetic field. A brief summary is provided at the end of the article.

\section{A method to measure the strength of the 2175~\AA\ bump from A$_{NUV}$ and A$_K$}

The extinction curve provides important clues about the properties of dust in space and the grains size distribution (see, for instance the review by Draine 2003 and references therein). The shape of the extinction curve is often modelled {\bf using} to polynomial fits that extend from the far UV to the infrared using as free parameter $R_V = A_V/E(B-V)$. In this work, we use the widespread extinction model derived by Fitzpatrick \& Massa 2007 (hereafter FM07). 

The curve is well behaved from the 3~$\mu$m to 0.3~$\mu$m; it allows a simple parametrisation in the infrared
that extends smoothly into the optical range so,
\begin{equation}
\frac {A_K}{A_V} = 1+\frac{1}{R} k(K-V)
\end{equation}
\noindent
with $k(K-V) = (-0.83+0.63R)\lambda ^{-1.84} -R$ and $\lambda = 2.2~\mu$m  for the average extinction
law according to  FM07. Hence,
\begin{equation}
\frac {A_K}{A_V} = 1+\frac{1}{R}(-0.19-0.85R)
\end{equation}

This smooth trend is broken in the UV range by the presence of the UV bump; five coefficients are required for the
mathematical representation of the extinction law in the UV (see Eq.~3), two of them to fit the UV bump to a 
Lorentzian function of ($1/\lambda$). Following FM07, the extinction law in the NUV band is parametrised as,

\begin{equation}
k(\lambda - V)  = E(\lambda - V)/E(B-V)  = 
\end{equation}
\[
c_1 + c_2x +c_3 D(x,x_0,\gamma) 
\]

\noindent
with $x=1/\lambda (\mu m)$ and, 
\begin{equation}
D(x,x_0,\gamma) = \frac{x^2}{(x^2-x_0^2)^2+x^2\gamma ^2}
\end{equation}

\noindent
Parameters $c_1, c_2, c_3, x_0$ and $\gamma$ are constants that depend on the specific line of sight.
The area of the bump is given by $A_{\rm bump} = \pi c_3/(2 \gamma)$ and its maximum intensity
by $I_{\rm bump} = c_3/ \gamma ^2$. 

The width of the bump varies considerably from line of sight (LoS) to LoS
with an average value of $0.993 \mu$m$^{-1}$ but its central wavelength remains estable 
(Fitzpatrick \& Massa, 1986). 
Absorption by small graphite particles was first hypothesised, as the main source 
of this strong absorption however, variations in the graphite grain shape and size 
should produce variations both in $\gamma$ and $x_0$
(Draine \& Malhotra 1993). Since the molecular structure of PAHs molecules is 
very similar to a portion of a graphite sheet, PAHs are a natural extension to the 
graphite hypothesis and are nowadays considered the main source of the bump
(see Draine 2003 review for more details).

By definition,

\begin{equation}
\frac {A_{\rm NUV}}{A_V} = 1+\frac{1}{R} k(NUV-V)
\end{equation}
\noindent
where $k(NUV-V)$ is weighted over the GALEX NUV band. Hence,

\begin{equation}
k(NUV-V)  = 
\end{equation}
\[
\int _{x_1}^{x_2} g_{NUV}(x) (c_1 + c_2x +c_3 D(x,x_0,\gamma)) dx 
\]
\[
= c_1<g_{NUV}> + c_2<xg_{NUV}>+c_3<g_{NUV}(x) D(x,x_0,\gamma)> 
\]

\noindent
with $g_{NUV}(x)$ the normalised transmittance function of the GALEX NUV channel \footnote{\bf http://galexgi.gsfc.nasa.gov/docs/galex/Documents/},  $x_1 = 1680$~\AA\
and $x_2 = 3000$~\AA . Thus, by definition,
\begin{equation}
< g_{NUV}> = \int _{x_1}^{x_2} g_{NUV}(x) dx = 1 
\end{equation}
\noindent
and,
\begin{equation}
<xg_{NUV}> = \int _{x_1}^{x_2} x g_{NUV}(x) dx = 4.49\\
\end{equation}
\noindent
The third term in {\bf Eq.~6} depends on $\gamma$ since {\bf it influences $D(x,x_0,\gamma )$}.
We have evaluated this integral for the various $\gamma$ values in 
FM07; they correspond to different LoS and extinction laws in the Galaxy. 
We have found that the integral can be fitted to a linear function of $\gamma$ such that,
\begin{equation}
<g_{NUV}(x) D(x,x_0,\gamma)> = 
\end{equation}
\[
-(0.82 \pm 0.02) \gamma + (1.38\pm 0.02)
\]
\noindent
with rms=0.003. This tight correlation was expected given the functional 
form of $D(x,x_0,\gamma)$ in the GALEX band and the strength of the 
2175~\AA\ feature. Since the area of the 2175~\AA\ bump
also depends on $\gamma$,   $A_{\rm bump} = \pi c_3/ 2 \gamma $, then,
\begin{equation}
<g_{NUV}(x) D(x,x_0,\gamma)> = -1.29 A_{\rm bump}^{-1}c_3^{2} + 1.38  
\end{equation}
\noindent

and therefore,
\begin{equation}
\frac {A_{NUV}}{A_K} = \frac {R+c_1+4.49c_2+c_3(-1.29 A_{\rm bump}^{-1}c_3^{2} + 1.38)} {0.15R-0.19}
\end{equation}

FM07 made a detailed evaluation of the extinction law parameters for several LoSs
(see Table 4 in their article) and using them, we have computed the $A_{NUV}/A_V$, $A_K/A_V $ 
and $A_{NUV}/A_K$ values for these specific LoSs; also, we have computed them for the average ISM coefficients\footnote{Parameters values for the average galactic extinction curve are: $x_0=4.592~\mu$m$^{-1}$, $\gamma = 0.922~\mu$m$^{-1}$, $c_1 = -0.175$, $c_2 = 0.807$ and $c_3 = 2.991$.} 
as derived by FM07 (see Table~1).
 
In general, our R(NUV-K) is slightly larger than the values computed by Yuan et al. (2013), 
R(NUV-K) = 6.75, as well as their evaluation from the FM07 extinction law $R(NUV-K) = 6.31$.

\begin{table}
\caption{The standard extinction law in the GALEX NUV band}
\begin{tabular}{llllll}
\hline
LoS   &  R$^{(a)}$ & $\frac{A_{NUV}}{A_V}$  & $\frac{A_K}{A_V}$  & $R(NUV-K)^{(b)}$ & $\frac {A_{NUV}} {A_K}$ \\
\hline
       HD698  &       3.94  &      2.56   &  0.09829  &       9.7 &      26.05 \\
      HD3191  &      2.81   &    2.923    & 0.07843   &    7.993  &     37.26 \\
   BD+57 245  &      2.97   &    2.828    &  0.08216  &     8.156 &      34.42 \\
   BD+57 252  &      2.97   &    2.889    &  0.08216  &     8.335 &      35.16 \\
  NGC457 P34  &      2.94   &    2.806    &  0.08149  &     8.011 &      34.44 \\
  NGC457 P13  &      3.11   &    2.588    &  0.08511  &     7.784 &      30.41 \\
   NGC457 P9  &      2.76   &    3.089    & 0.07718   &    8.314  &     40.03 \\
 ISM average  &     3.001   &    2.735    &  0.08284  &     7.958 &      33.01 \\
\hline
\end{tabular}
\begin{tabular}{ll}
$^{(a)}$ & $R = A_V/E(B-V)$ \\
$^{(b)}$ & $R(NUV-K) = (A_{\rm NUV}-A_K)/E(B-V)$ \\
\end{tabular}
\end{table}

\noindent

$A_{NUV}/A_K$ and $A_{\rm bump}$ have been evaluated for all the LoS in FM07 (see Table~1) and plotted in Fig.~1.
Notice that $A_{NUV}/A_K$ and $A_{\rm bump}$ values are well correlated unless for HD~3191, which seems to be a peculiar
\footnote{HD~3191 also affects significantly the average ISM law, as parametrized by FM07}.
For this reason, it  has been excluded, {\bf and the resulting fit in Fig.~1 is:}
\begin{equation}
A_{bump} = (0.106 \pm 0.008)\frac {A_{NUV}}{A_K}+ (2.0 \pm 0.3)
\end{equation}
\noindent

This result is not a surprising; it just points out that the ISM extinction law admits a rather simple parametrisation
from the near infrared to the near ultraviolet unless for the strength of the bump. Therefore, for the dust grain size 
distributions observed in the ISM, the main deviations from this trend are caused by the UV bump.   

Likewise, the strength of the bump can be evaluated as a linear function of the NUV to K band extinction for the ISM.
Note that this {\it semi-empirical relation} has been derived using the LoSs in FM07 that sample a broad range of 
extinction laws ({\it i.e.}, R-values and bump strengths) and it is independent of the physical origin of the bump
be it PAHs, small graphite grains or any other possible source ({\it e.g.}  fullerens, see Wada et al. 1999). 

In principle, similar expressions could be derived for other bands however, the availability of the near infrared extinction map of the TMC 
(LLA10) makes the $A_{NUV}/A_K$ rate, well suited for our purpose. Note that emission from average warm dust grains
dominate the K-band while absorption by PAHs or alike dominates the NUV band.

Mid-infrared images have also been used to test dust evolution by comparing the emission from PAHs and very small grains with the 
radiation from large dust grains (see {\it i.e.} Flagey et al. 2010). However, the NUV range is ideally suited for this
purpose given its high sensitivity to small particles; the estimated oscillator strength per nucleon is $n_Xf_X/n_H \simeq 9.3 \times 10^{-6}$ for the 2175~\AA\ bump (Draine, 1989).

\begin{figure}
\begin{center}
\includegraphics[width=8cm]{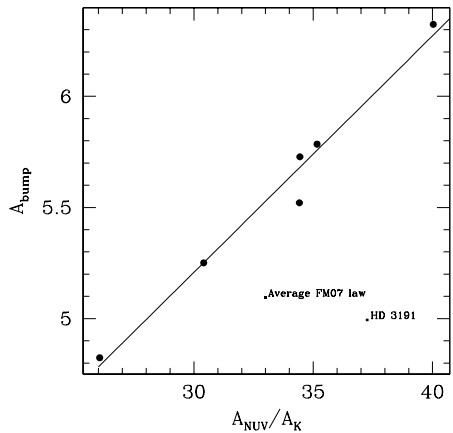} \\
\end{center}
\caption{Relation between the area of the 2175~\AA\ bump and the relative
NUV versus K extinction for the FM07 extinction law.}
\label{bump}
\end{figure}

\begin{figure}
\begin{center}
\includegraphics[width=8cm]{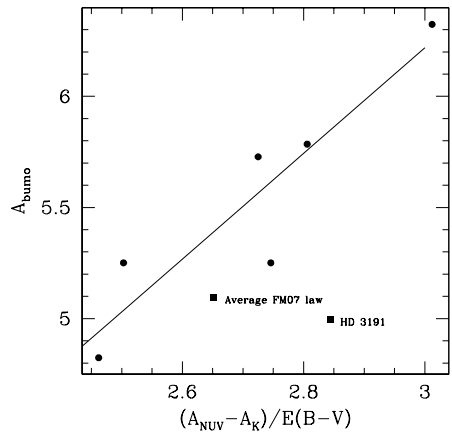} \\
\end{center}
\caption{Relation between the area of the 2175~\AA\ bump and the $(A_{NUV}-A_K)/E(B-V) $
for the FM07 extinction law.}
\label{bump}
\end{figure}

For completeness, we also note that,
\[
A_{NUV}-A_K = E(B-V) (0.19 + c_1 +4.49 c_2 -1.29 A_{\rm bump}^{-1}c_3^{3}
\]
\begin{equation}
+1.38 c_3 +0.85 R)
\end{equation}
and thus, $(A_{NUV}-A_K)/ E(B-V)$ can be parametrised in terms of the strength of the bump and, as above, a simple scaling can be made in terms of the extinction law for various LoS,
\begin{equation}
A_{\rm bump} = (2.4 \pm 0.5) \frac {A_{NUV} -A_K} {E(B-V)} - (0.9 \pm 1.5)
\end{equation}
\noindent
though the fit is significantly less accurate (see Fig.~2).

\section{The GALEX survey of the TMC}

The baseline of the GALEX All Sky Survey was completed in 2007.  GALEX AIS covers 26,000~deg$^2$ ($\sim 63$\% of the Sky) and provides broadband imaging in two UV bands. GALEX obtained 197 images on the Taurus molecular cloud with a total coverage of $\sim 200$ deg$^2$ 
(see Fig.~3); note that the GALEX field of view is circular with radius 0.6$ ^{\rm o}$. The survey avoids the central region of the TMC where TMC-1 is located, though some pointings are made  around TMC-2. Some molecular cores like L1495, L1498, L1544, L1515, L1548 and L1552 are partially covered 
in the survey. 

The GALEX mission provides as output products for each tile (or image in the AIS survey) several files containing the pipeline processed images in the FUV and NUV bands, the intermediate calibration files and the catalogue of sources identified in each pointing; the sources in the catalogue are identified with the SExtractor procedure (Morrissey et al. 2007) and their FUV and NUV magnitudes are provided. Typically, the number of sources in the catalogue outnumbers by a factor of 2-3, the number of sources that can be identified as such from a simple inspection of the images. The detection procedure clearly suffers from overestimation of actual sources. The number of spurious detections is very large. 

As the TMC is close to the galactic plane, a cross-identification with galactic sources from the Fourth USNO CCD Astrograph Catalog - U.S. Naval Observatory (UCAC4; Zacharias et al. 2013) and the 2MASS surveys (Strutskie et al. 2006) has been carried out. Typically, for each of our GALEX fields, there are a factor of 3 more sources in the 2MASS survey and a factor of 10 in the UCAC4 survey; thus, they outnumber by far the UV sources identified by the GALEX pipeline and can be used to check the reliability of the GALEX identification.  To complete the cross-identification of our sample with sources at different wavelengths, we have also cross-correlated our list with the WISE all-sky catalogue (Cutri et al. 2012) using the same criteria than for the other two databases.

Magnitude completeness limits for 2MASS are 15.8, 15.1 and 14.3 mag in J, H and K$_\mathrm{s}$
bands, respectively. These limits are well above the magnitude of M-type stars at the distance of
the Taurus-Auriga molecular complex. J magnitude for a 4~Myr old M5 star is $\sim 5.5$ (e.g. Siess et al. 2000). 
At a distance of 140 pc, the magnitude of such star $J \sim 11.5$~mag. Even in the case of notably ISM extinction 
$A_\mathrm{J} = 4$~mag, the star would be detected by 2MASS at the distance of the molecular cloud. 
Similarly, UCAC4 is complete from the brightest stars to about $R = 16$~mag. This means that 
UCAC4 is complete for mid-M stars at the distance of the Taurus cloud for extinctions
$A_\mathrm{V} < 2$~mag. As we do not expect large extinctions for our stars, because they were 
selected from the UV, we can conclude that we are complete for members of the star-forming 
region for the entire luminosity function, from the more massive to the less massive stars. 
Completeness limit for a typical 100s exposure time GALEX/AIS image in NUV and FUV is 
$\sim 20$~mag in both cases (Bianchi et al. 2013).

{\it Bona fide} GALEX sources have been identified by having a 2MASS counterpart within a search radius of 3~arcsec. This search radius was carefully selected after a precise study of the shift between 2MASS and GALEX sources in the TMC (see G\'omez de Castro et al 2011 for details). After the cross-correlation, we kept a total of 163,313 UV sources as reliable detections. All them have been detected in the NUV band but only 10\% of the sources have a FUV counterpart. Since this research is based on stellar statistics, we worked with the sample of confirmed NUV sources (163,313 sources). Only 31 TTSs and and 63 candidates to TTSs are found in the area covered by this survey (G\'omez de Castro et al. 2015) thus, their impact in the overall statistics is minimum. 
 
\begin{figure}
\includegraphics[width=8cm]{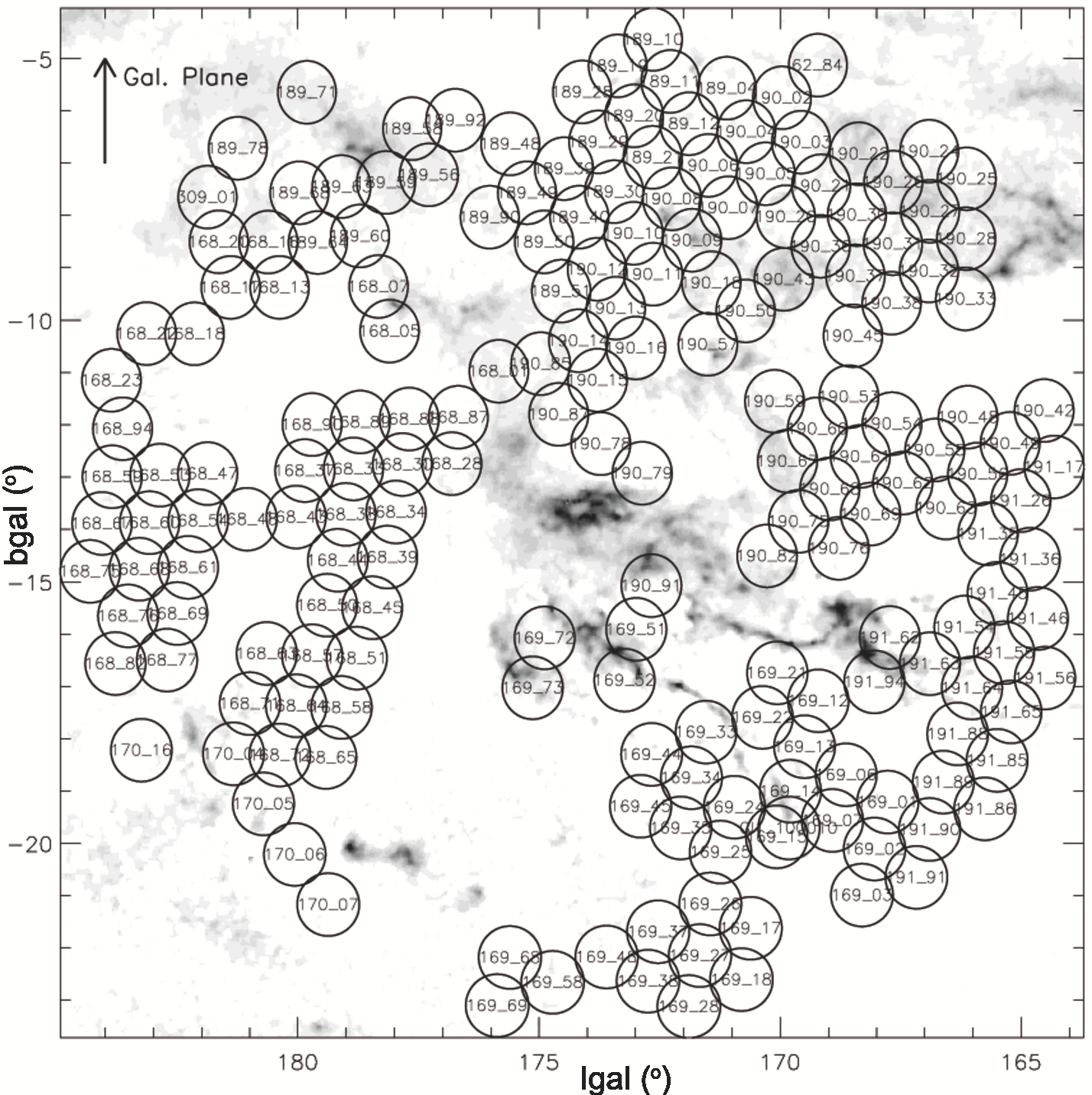}
\caption{Fields observed by GALEX (AIS GR5) towards the  Taurus Molecular Complex (open circles) 
are overlaid  on the map of hydrogen column density ($N_H$) produced by LLA10
from 2MASS star counts. }
\label{ais}
\end{figure}

In Fig.~4, the surface density of stars towards the TMC is plotted; the location of the main molecular cores as well as the brightest UV sources in the field (from Lee et al. 2006)  are marked. 
The NUV surface density of stars is overlaid on the large scale extinction map of the area obtained by
Lombardi et al (2010) from the 2MASS survey of the TMC.  A comparison between both maps (see also Fig.~4 in LLA10), points out that most of the area surveyed by GALEX has $A_K \leq 0.2$~mag that roughly corresponds\footnote{The average ISM extinction law, as described in 
Sect.~2, has been used for this scaling.} to $A_V \leq 2.2$~mag (or $N_H \leq 4 \times 10^{21}$~atoms~cm$^{-2}$). The density of stars drops to zero
at higher gas columns. For instance, the extension of some well known filaments, such as L1495 or L1498,  shows in the stellar density map because
the density of stars drops to 0. Also, some diffuse structures are noticeable. For instance,
at $b_{\rm gal} = -13^{\rm o}$, the stellar density varies  from  5-6 stars per 3~arcmin$^2$ at $l_{\rm gal} = 165 ^{\rm o}$ (westwards of the TMC) to 2-3 stars per 3~arcmin$^2$ at $l_{\rm gal} = 180 ^{\rm o}$; this feature passed unattended in previous CO maps of the area.

\begin{figure}
 \includegraphics[width=8cm]{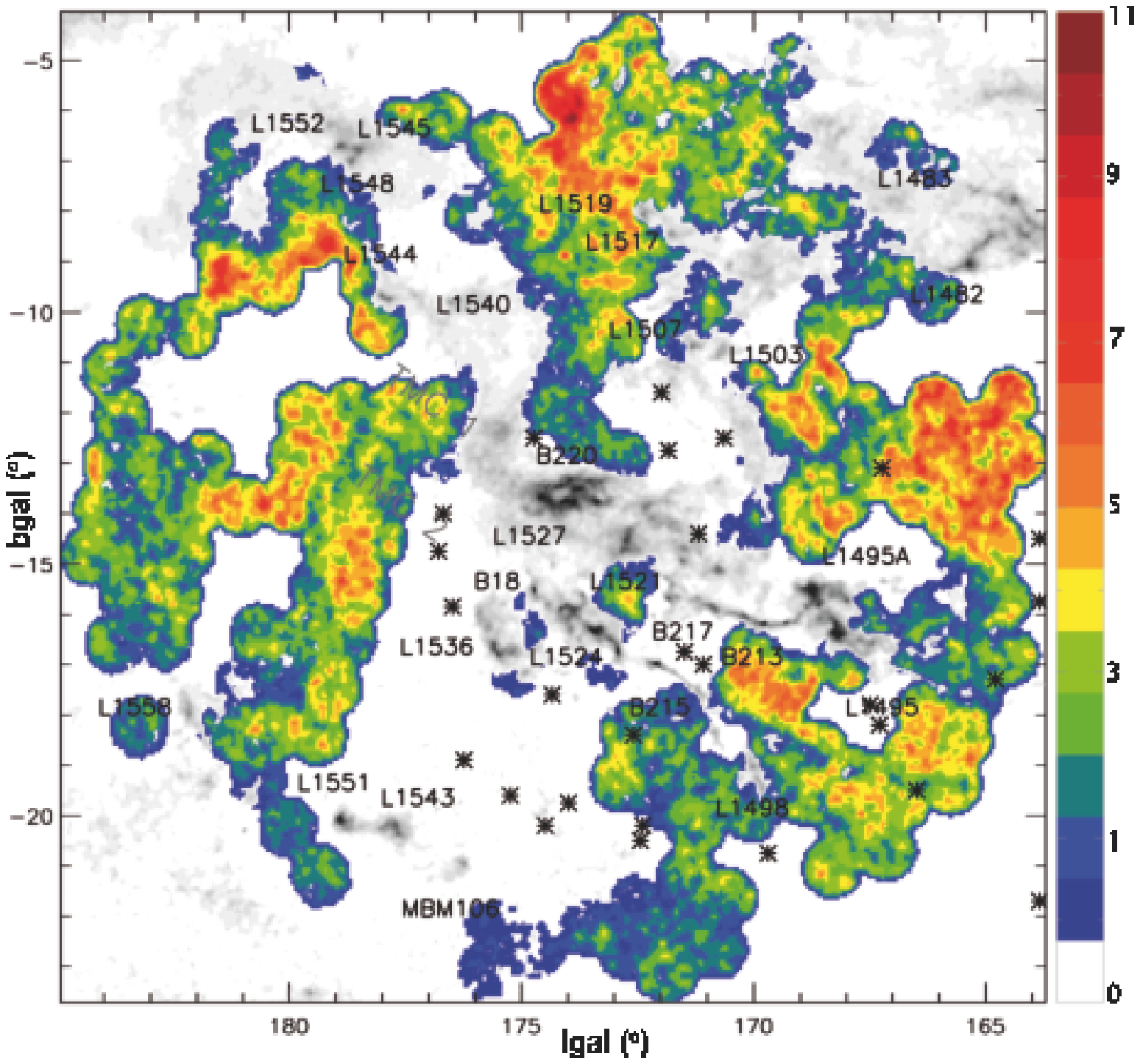} \\
\caption{Density of NUV GALEX sources in the TMC. The densities are colour coded in stars per 3 arcmin$^2$ (see lateral bar). The 2MASS extinction map (LLA10) is overlaid for reference. The location of the main molecular cores and the main UV sources in the field (asterisks) are marked. }
\label{NUV_clean}
\end{figure}

\section{Methods}
\subsection{The Wolf method}

The Wolf method is designed to measure extinction using as reference for the stellar luminosity function a nearby stellar field, assumed to be unextincted. The luminosity function is defined in terms of the cumulative star counts distribution per apparent magnitude bin, in the fiducial field (Wolf 1923: see also  Gorbikov \& Brosch, 2011
for a recent application of the method to study the extinction law towards the north celestial cap).
The method is based on the fact that all the stars behind the same absorbing cloud are extincted by the  same amount, $A_{\rm \lambda}$. As a result,  the cumulative star counts distribution as a function of the apparent magnitude, $N(m(\lambda))$, is homogeneously shifted to lower apparent magnitudes in any dusty area. The shift of the cumulative counts distribution is a direct measure of the relative extinction between the fiducial field and any given field. The process is illustrated in Fig.~5 for tile AIS190\_sg57. A control tile, tile  AIS189\_28, has been used for calibration purposes since 
extinction is not apparent neither in the UV data (see Fig.~3) nor in the infrared data (see LLA10).  
The NUV cumulative distribution in AIS190\_sg297 is clearly  shifted with respect to the reference field pointing out a NUV extinction of 
$A_{\rm NUV} = 2.8$ mag. 

\begin{figure}
\includegraphics[width=8cm]{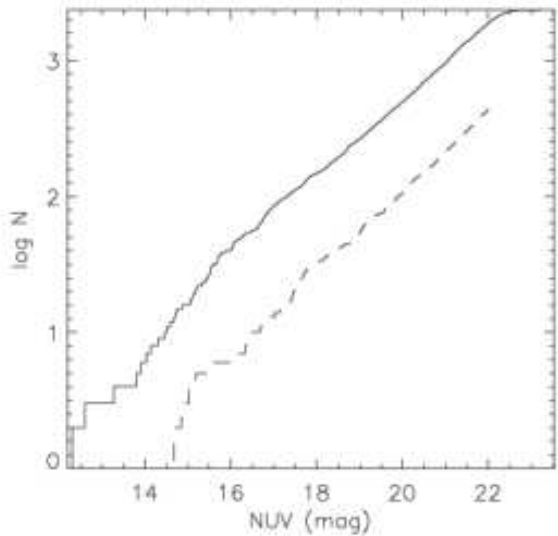}
\caption{Determination of the extinction using the Wolf method for tile AIS190\_sg27. The plot illustrates the cumulative counts  
distribution in AIS190\_sg27 (solid line) compared with  AIS189\_28, the reference tile (dashed).}
\label{wolf}
\end{figure}

Following this procedure both $A_{NUV}$ and  $A_K$ values have been computed however, the $A_{\rm NUV}/A_K$ 
ratios obtained are unusually small and inconsistent with the interstellar extinction law ({\it e.g.} FM07).
These small ratios are obtained because the GALEX AIS fields contain areas with and without dust clouds; 
this geometric dilution together with the clumpiness of the ISM produces a weighted average of the 
$A_{\rm NUV}/A_K$ value over the GALEX AIS field of view (see Appendix A for more details). Therefore, 
it is critical to work with spatial resolutions that adapt well to the characteristic scales of the ISM 
structures to be studied such as filaments, globules etc…, typically about 0.1-0.4~pc.

Unfortunately, the Wolf method requires a good statistics to sample properly  the luminosity function 
forcing the resolution element to be much larger than 0.1-0.4~pc (2.5-10 arcmin scales at the
TMC distance of 140 pc). A fine tuning between resolution and statistics is required.
The star counts method is better suited for this purpose.

\subsection{Star counts}

Since the seminal work  by van~Rhijn (1929), star counts have been used to measure extinction in galactic fields. A detailed description of the  method is given by Bok \& Cordwell (1973).  This technique has been  extensively used in astronomy and has been successfully applied to measure  the gas column to Taurus  using the R plate of the Palomar Observatory Sky Survey (Cernicharo \& Bachiller, 1984). 

Basically, the apparent magnitude of a given star in the NUV band, $m(NUV)$, is given by:
\begin{equation}
m(NUV) - M(NUV) = -5 + 5 \log \frac{d}{\rm pc} + A_{\rm NUV}
\end{equation}
\noindent
where $M(NUV)$ is the absolute magnitude, $d$ is the distance and $A_{\rm NUV}$ is the extinction in the
NUV band caused by the intervening gas. In its most usual application, star counts in a given field are
compared with the predictions for an unextincted, nearby field and the extinction is given by,
\begin{equation}
A_{\lambda}=\log(N^{*}_{\lambda}/N_{\lambda})/b_{\lambda}
\end{equation}
\noindent
where $N_{\lambda}$ are the observed counts, $N^{*}_{\lambda}$ are the expected counts from a non-extinguished field, and $b_{\lambda} = d\log N(m_{\lambda})/dm_{\lambda}$ 
(with $m_{\lambda}$ the apparent magnitude at wavelength $\lambda$) is a measure of the 
slope of the luminosity function in the area under study. 

The NUV luminosity function has not been determined for the Galaxy thus we have derived it in this work for the TMC region.
Firstly, we have determined it for the reference field (AIS189\_28), which we consider unextincted and has 2,340 sources. 
We have calculated it to be well fitted by,

\[
\log N(m_{NUV}) = (0.294 \pm 0.002) m_{NUV} -2.55 \pm 0.04
\]
with $RMS = 0.046$ and thus, $b_{NUV} = 0.294$ (see Fig.~6).  We also have tested whether there are significant variations
over the TMC field. As shown in the bottom panel of Fig.~6, the slope remains constant (within the error bars) pointing out that
only minor variations are detected over the field, if any.  

\begin{figure}
\includegraphics[width=16cm]{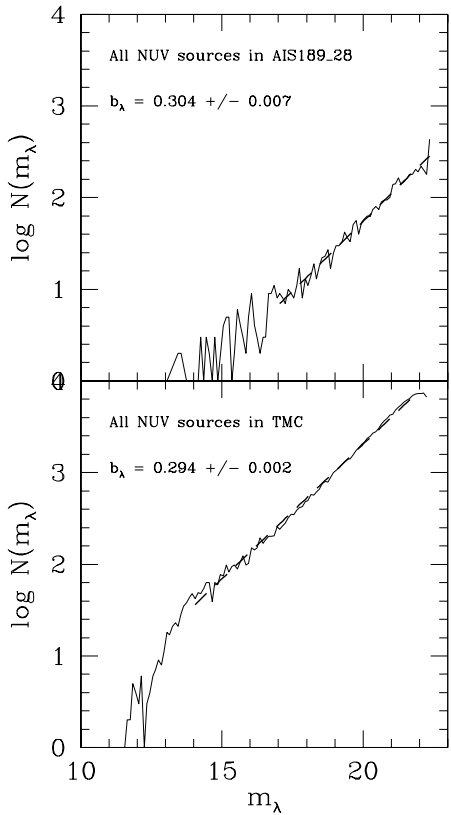}\\
\caption{Stellar luminosity function in the TMC in the NUV band. {\bf Top:} NUV luminosity function in the reference field, {\bf bottom:}
luminosity function derived from all NUV sources in the AIS TMC survey. The best fit to the slope is shown by 
a dashed line on the fitted region.  }
\label{Lumfunc}
\end{figure}

The accuracy of  the $A_{NUV}$ values derived from the star counts method is (see Dickman 1978),
\begin{equation}
\frac {\delta NUV}{NUV} =   (\log \frac{N^*_{NUV}} {N_{NUV}} )^{-1} (\frac {N_{NUV}+N^*_{NUV}}{N_{NUV}N^*_{NUV}})^{1/2}
\end{equation}
with,
\begin{equation}
\delta A_{NUV} = \frac{1}{b_{NUV}}(\frac  {N_{NUV}+N^*_{NUV}} {N_{NUV}N^*_{NUV}} )^{1/2}
\end{equation}
Thus, for $A_{NUV}=0$ (or $N^*_{NUV} = N_{NUV}$), $\delta A_{NUV} = 0.1$~mag.

The maximum $A_{NUV}$ is observed in tile AIS190\_43 with $N_{NUV} = 136$, and thus $A_{NUV} = 3.2$~mag 
with $\delta A_{NUV} = 0.3$~mag.

If the statistics {\bf are good then both the Wolf and star-counts should produce  very similar results.} 
Average $A_{\rm NUV}$’s have been calculated by both methods for all the GALEX fields and,
as shown in Fig.~7, the only discrepancy is at high extinctions (low star counts) since the   
Wolf method produces slightly higher extinctions than the star-counts method. 
The regression line is:
\[
A_{\rm NUV}(\rm Wolf) = (1.10 \pm 0.03) A_{\rm NUV}(\rm Counts)+ (0.08 \pm 0.04)
\]
\noindent
with $rms = 0.27$. 

\begin{figure}
\includegraphics[width=8cm]{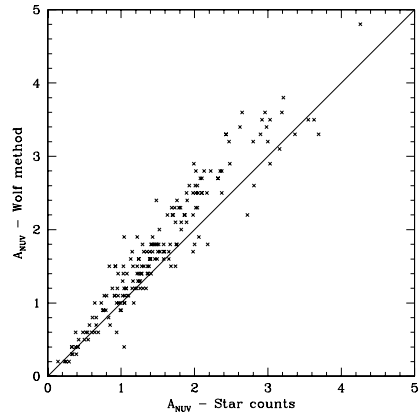}
\caption{Comparison between the A$_{NUV}$ values obtained by direct star counts and by using the Wolf method. The
straighline marking A$_{NUV}$ - Wolf method equal to A$_{NUV}$ -Star counts is plotted for comparison.}
\label{wolftest}
\end{figure}

\subsection{Galactic latitude correction}

Stellar density declines significantly with galactic latitude\footnote{As pointed out in the text, we have not found significant variations in the luminosity function, as a simple inspection of Fig.~6 shows.}. To correct for this effect, we have computed the variation of $N^*_{NUV}$ with galactic latitude and derive a correction factor, $dA$, such that, 

\begin{equation}
A^0_{\rm NUV} = A_{\rm NUV}-dA_{\rm NUV}
\end{equation}
\noindent
being  $A_{\rm NUV}$, the extinction determined with the star counts method, Eq.~16, and  $A^0_{\rm NUV}$, the final value of extinction. To evaluate this correction, we have made use of the Besancon model of the Galaxy (Robin et al. 2003; Czekaj et al. 2014, see also URL: model.obs-besancon.fr), and the $A_K$ map of the TMC evaluated by LLA10 (see details in Appendix B). The correction 
factor is found to be,
\begin{equation}
dA = (0.0640 \pm 0.0005) b_{gal}(^o) - (0.38 \pm 0.01)
\end{equation}
\noindent
(with $rms = 0.025$).

The variation of the stellar luminosity function between $l_{\rm gal} =165 ^{\rm o}$ and  $l_{\rm gal} = 180  ^{\rm o}$ is negligible according to the Becanson standard model.

\section{Results: Variation of $A_{NUV}/A_K$ over the TMC}

Cores and filaments in molecular clouds have typical physical scales of
0.1-0.4 pc that correspond to 2.5-10 arcmin at the TMC distance (140 pc). Unfortunately,  the UV
statistics is too poor at resolutions of 3~arcmin$^2$ (see Fig.~4).
Thus, we have followed an incremental approach using two intermediate bin sizes of 24~arcmin$^2$ and 12~arcmin$^2$. 

NUV extinctions have been calculated by applying the star counts method (see Sect.~4) including the galactic latitude correction
derived in Appendix B. To compute K-extinctions we have resourced directly to Froebrich et al. (2007) map.
The map has been rebinned and re-scaled to K-magnitude; note that  F07 provide the electronic version of the map (FITS format) 
in visual extinctions ($A_V$), but these $A_V$ values are obtained by the authors after applying a simple scaling from infrared to 
visual magnitudes assuming that the standard ISM extinction law holds (Draine 1989). 

As shown in Fig.~8, there is a significant variation of $A_{NUV}/A_{K}$ across the region. 
$A_{NUV}/A_{K} \simeq 33$ in the diffuse ISM, around the TMC in agreement with
FM07. However, westwards, at the tail of the cloud, where the main filaments reside, and the column density of
molecular gas increases, $A_{NUV}/A_K$ decreases pointing out the weakening of the UV bump and, in general, 
a larger average size of the dust grains when compared with the  diffuse ISM. This trend is confirmed at 
both intermediate resolutions.

Some of the areas mapped by GALEX have very low and thus, uncertain, $A_K$ values (LLA10 estimates $dA_K = \pm 0.04$~mag).
This results in abnormally high $A_{NUV}/A_K$ rates, especially at high galactic latitudes (see Fig.~8).

Experimentally, the UV bump is known to be produced by mixtures of sufficiently large, neutral PAHs (Steglich et al. 2011).
Moreover, the UV-visible measurements provide a reliable fingerprint of the presence of specific PAHs since the electronic 
transitions of these species are detected in this range while the near infrared vibrational
bands are not molecule specific. 

A recent UV-visible (3050-3850~\AA\ ) study  based on the observations towards five lines of sight with moderate reddenings 
measured PAHs abundances  two orders of magnitude smaller than those in the diffuse ISM (Gredel et al. 2011);
the measurements were made over column densities of  $N_H = 5.7 - 9.2 \times 10^{21}$~cm$^{-2}$, 
slightly above the dominant column density of our study in the TMC, $N_H \leq 4 \times 10^{21}$~atoms~cm$^{-2}$.
This study was sensitive to molecules such as anthracene, pyrene or benzofluorene. Laboratory experiments suggest that 
the UV bump is produced by larger PAHs with sizes above 50-60 carbon molecules (Steglich et al., 2010)
thus, our results suggests that large PAHs are neither abundant in traslucent clouds.

$A_{\rm NUV}/A_K$ is small close to the filamentary networks (compare Fig~3 with Fig~8) where also mid infrared data
have detected the formation of large dust grains (Flagey et al. 2010), as high density of molecular gas makes 
collisional processes more efficient assisting grain growth. In addition, the large density of dust in the filaments 
effectively shields the growing mantles against the environmental UV radiation.
A detailed comparison between Figure 4 and 8, shows that the bump is less prominent towards the tail of the 
filaments leading to L1495 and L1498. The first results from the Herschel survey of the Gould Belt in Taurus, 
show  a fraction of this area: N1-3 for the filament leading to L1495 and S1-4 for that leading to L1498. 
The analysis of Herschel data shows that these filaments consist really in a  hierarchy of cores of molecular gas with sizes in the range $0.024 – 2.7$pc that covers the full core mass function from unbound clumps to (gravitationally bound) pre-stellar cores, being all part of the same population (Kirk et al. 2013). Our detection of increased average size of dust grains around this area is consistent with 
the enhancement of the molecular gas and dust density that drives the formation of molecular cores. 

In the global scenario developed by GdCP92, matter in the TMC is being gathered by the local action
of the Parker-Jeans instability. Matter lifted to the halo by the action of the instability,
rains back along the magnetic field lines reaching velocities comparable to the Alfv\'en speed
in the undisturbed gas layer in the galactic plane. These velocities agree well with the
clouds and pre-main sequence stars velocities that roughly run eastwards (see GdCP92). 
Fresh material is thus expected to be gathered at the rear of the cloud,
where it is, indeed. As the density increases with respect to the average
ISM values, so does the opacity protecting dust grains from the ambient UV radiation field
and favouring grain growth. In this sense,
our results are consistent with the expectations. Unfortunately, the long wavelength Alfv\'en wave
mapped by Moneti et al. (1984) was not properly surveyed by GALEX.

\begin{figure}
\includegraphics[width=7cm]{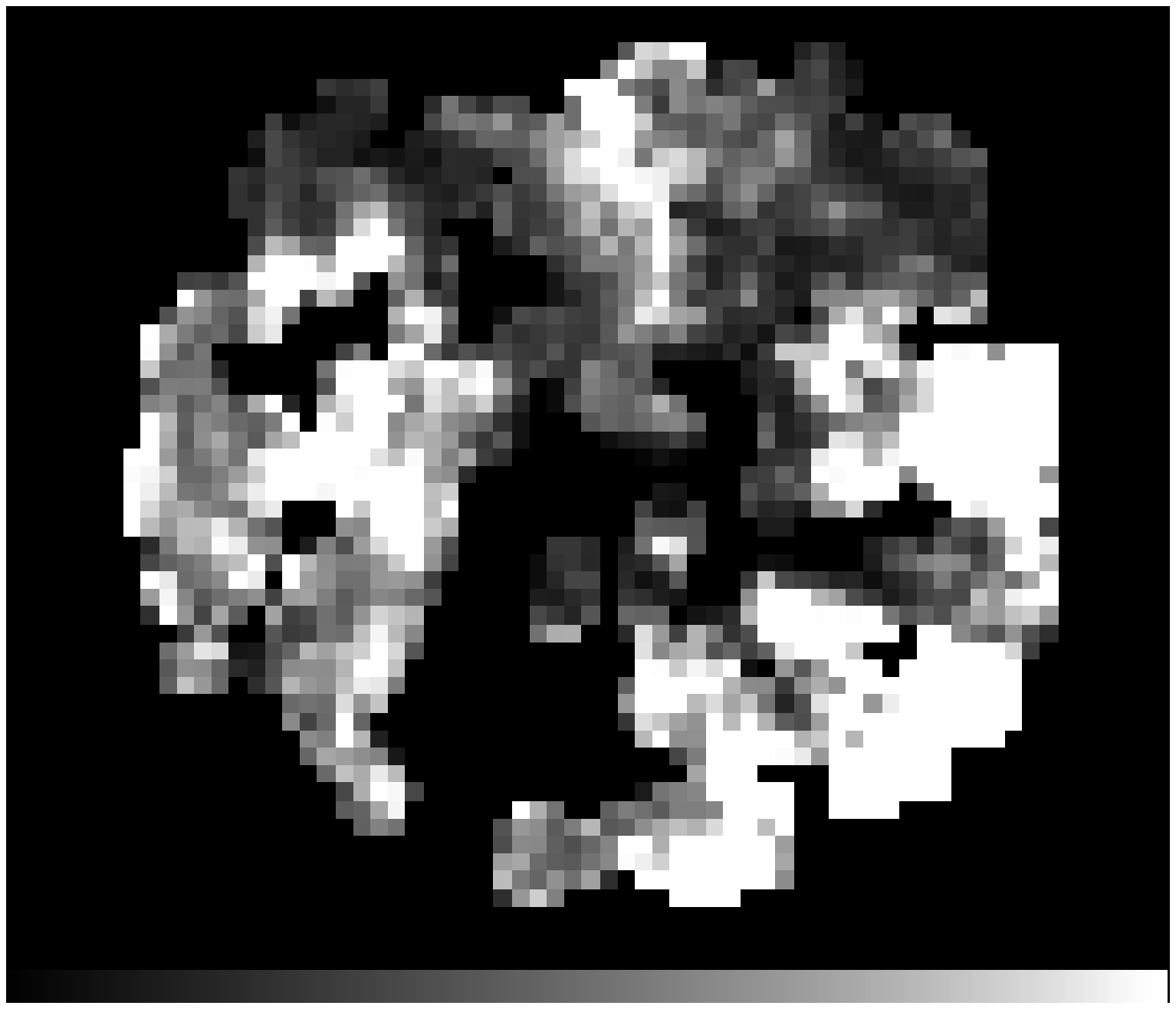} \\
\includegraphics[width=7cm]{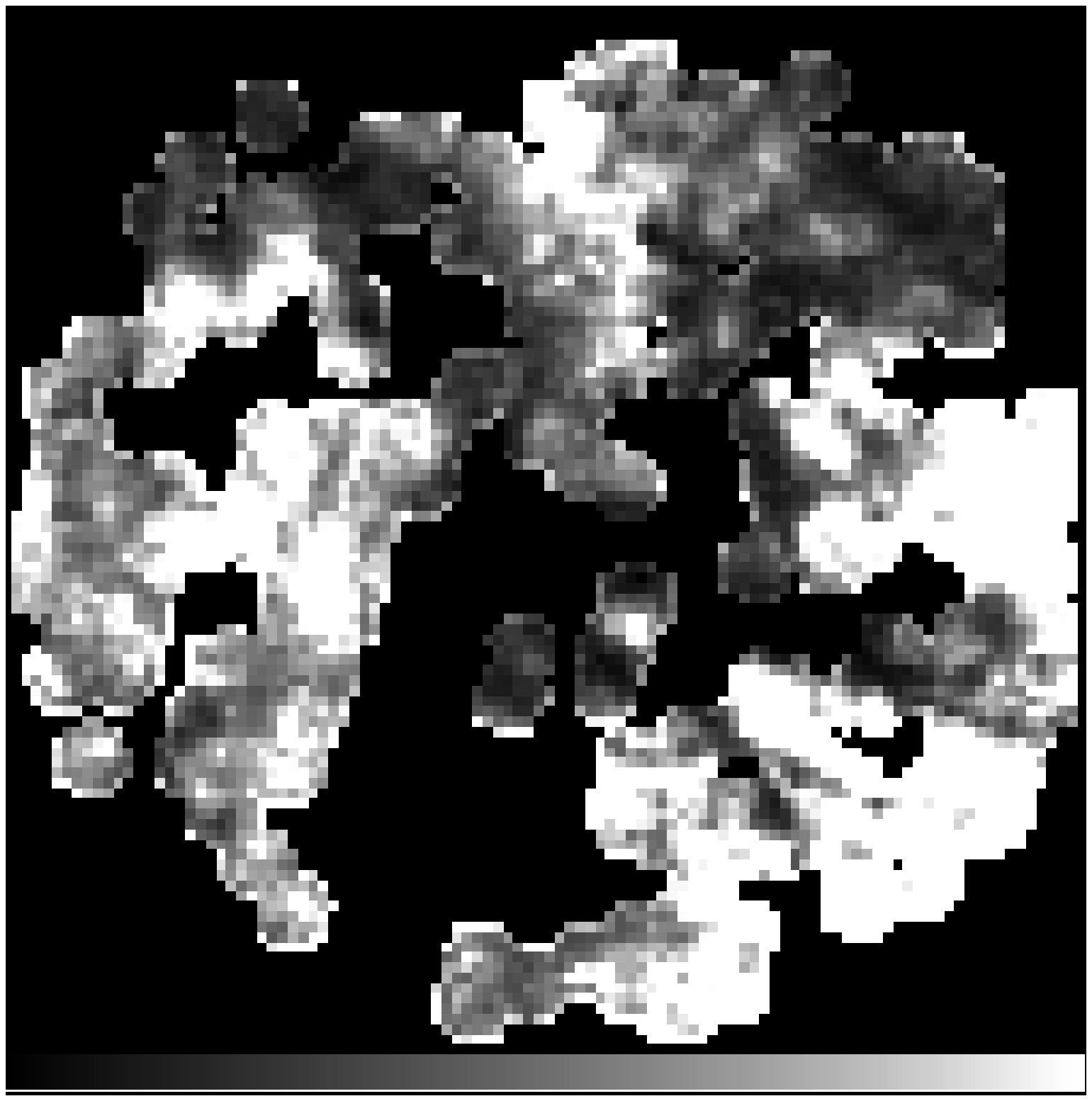}
\caption{$A_{NUV}/A_K$ on the TMC computed for bins of 24~arcmin$^2$ and 12~arcmin$^2$, top and bottom panels respectively. 
The gray scale runs  from $A_{NUV}/A_K = 0$ (black) to  $A_{NUV}/A_V \geq 33$ (white).
$A_{NUV}/A_K \geq 33$ corresponds to the predictions of the standard ISM extinction law (see Table~1).}
\label{anuvak}
\end{figure}

\section{Conclusions}
In this work, we have shown the  potentials of wide field UV imaging to study the distribution and properties of diffuse dust clouds in space. Our main contributions to the analysis of the UV data have been:

\begin{itemize}

\item to show that the wide field UV images obtained by means of the GALEX AIS are able to detect very low dust columns that are not detected through infrared surveys (2MASS).  These images are obtained with a rather modest telescope (40~cm) in 150~s (barely 2.5 minutes), in average.
\item the stars luminosity function has been derived for the GALEX NUV band in the galactic anticenter region. $b_{NUV}$ has been computed. This value is suitable to be compared with measurements across the Galaxy.  
\item a generic method to determine the relative abundance of small dust grains from star counts and surveys has been described. 

\end{itemize}

These developments have been used to detect variations in the extinction law  across the TMC. The extinction law tends to be greyer in
the rear of the cloud and in the networks of filaments, than on the clean ISM. This is consistent with the
generic picture of large dust grains being formed in molecular clouds and small dust grains being more abundant in the ISM.
It is worth remarking that the areas where dust is being gathered are at the rear of the cloud, as expected
if star formation at the head of the TMC is associated with a local storage of mass being gathered by the action of 
the Parker-Jeans instability. 

Unfortunately, the depth of the GALEX~AIS is not enough to have good statistics 
in the areas around the filaments so, we cannot test effects like grain growth.
Note that according to,
\[
A_{NUV}^{lim} \simeq \log(N^{*}_{NUV}/N_{NUV}^{lim})/0.29
\]
and setting $N_{NUV}^{lim}=1$, $ A_{NUV}^{lim}$ depends on the number of sources
in the test field (which is the maximum) and thus, on the resolution. To reach
spatial resolutions of 0.05 pc, reasonable to study the filaments structure,
and $ A_{NUV}^{lim} = 4$, it is required that $N^{*}_{NUV} = 16$ per 1.2~arcmin$^2$, 
which represents a factor of 14 increase in stellar density (stars/arcmin$^2$)
with respect to GALEX~AIS. The limiting magnitude should rise from $NUV = 22.3$ mag 
(see G\'omez de Castro et al. 2015)
to $ NUV = 23.5$~mag.

\section*{acknowledgments}
This work has been partially funded by the Ministry of Science and Innovation of Spain through grant: AYA2011--29754-C03-C01 and AYA2011--29754-C03-C03. This article is based on data obtained by the  NASA mission Galactic Evolution Explorer (GALEX).

\appendix

\appendix

\section{Estimate of the impact on the $A_{NUV}/A_K$ rate of the filling factor of the dust clouds within the GALEX AIS field of view}

Let  $A_{NUV}$ and $A_K$ be the extinctions evaluated from the star counts and affected by the filling factor of the extincted over un-extincted areas in the field of view.
Let  $A_{NUV}^f $ and $A_K^f$ the true, average values, taking into account only the extincted areas.
Then,
\begin{equation}
\log \frac {N_{NUV,max}}{N_{NUV}}  =  b_{NUV} A_{NUV}^f 
\end{equation}
\begin{equation}
\log \frac {N_{K,max}}{N_{K}}  =  b_K A_{K}^f 
\end{equation}

\noindent
and, 
\begin{equation}
\delta A_{NUV}  =  \frac {1} {b_{NUV}} \log \chi =  A_{NUV}^f - A_{NUV}
\end{equation}
\begin{equation}
\delta A_K        = \frac {1} {b_K}       \log \chi  = A_K^f - A_K 
\end{equation}
\noindent
with  $\delta A_{NUV}$ and $\delta A_K$ the excess extinctions caused by the geometric filling factor of dusty areas in the 
GALEX field of view, $\chi$.  With this definition, 

\begin{equation}
\frac {A_{NUV}} {A_K}  =  \frac {b_K} {b_{NUV}} \frac {b_{NUV} A_{NUV}^f - \log \chi} {b_K A_{K}^f-\log \chi} 
\end{equation}
\noindent
and after some algebra,

\begin{equation}
\frac {1}{b_{NUV}} \log \chi =  A_{NUV}^f  \frac {\frac {A_{NUV}}{A_K} \frac {A_K^f}{A_{NUV}^f}-1} {\frac {b_{NUV}}{b_K} \frac {A_{NUV}}{A_K} -1} 
\end{equation}
\begin{equation}
\frac{1}{b_{K}} \log \chi =  A_{K}^f  \frac {\frac {A_{NUV}}{A_K} -\frac {A_{NUV}^f}{A_{K}^f}} {\frac {A_{NUV}}{A_K} -\frac {b_K}{b_{NUV}} } 
\end{equation}

\noindent
or,
\begin{equation}
\delta A_{NUV}  =  A_{NUV}^f \frac{\frac{A_{NUV}}{A_K} \frac{A_K^f}{A_{NUV}^f}-1} {\frac {b_{NUV}}{b_K} \frac {A_{NUV}}{A_K} -1} 
\end{equation}

\begin{equation}
\delta A_K  =  A_K^f  \frac{\frac{A_{NUV}}{A_K} -\frac{A_{NUV}^f}{A_{K}^f}} {\frac {A_{NUV}}{A_K} -\frac {b_K}{b_{NUV}} } 
\end{equation}
\noindent
making use of the $b_K$ and $b_{NUV}$ values determined from the luminosity functions (see Sect. 4.1) and assuming 
$A_{NUV}^f/A_K^f=33.01$, the average ISM value, we derive,
 
\begin{equation}
\frac{\delta A_{NUV}} {A_{NUV}^f} = \frac{\frac{A_{NUV}}{A_K} \frac{1}{33.01}-1} {0.94\frac {A_{NUV}}{A_K} -1} 
\end{equation}
\begin{equation}
\frac{\delta A_K}{A_K^f} =   \frac{\frac{A_{NUV}}{A_K} -33.01} {\frac {A_{NUV}}{A_K} -1.07}  
\end{equation}

Finally, note that $(A_{NUV}-A_K)$ is basically insensitive to the filling factor since,

\begin{equation}
A_{NUV} - A_K =  \frac{1}{b_{NUV}} \log \frac{N_{NUV,max}}{N_{NUV}} - \frac {1}{b_{NUV}} \log \chi 
\end{equation}

\begin{equation}
A_{NUV} - A_K =  \frac{1}{b_{K}} \log \frac{N_{K,max}}{N_{K}} - \frac{1}{b_{K}} \log \chi 
\end{equation}

\begin{equation}
A_{NUV} - A_K = A_{NUV}^f - A_K^f - ( \frac {1} {b_{NUV}} - \frac {1} {b_K}) \log \chi 
\end{equation}
assuming $b_{NUV} \simeq b_K$. Unfortunately, this colour has a softer dependence on the strength of the 2175~\AA\ bump
than the rate $A_{NUV}/A_K$ as shown in Sect.~2. 

This short exercise provides a first order estimate and neglects the effects caused by the cloud clumpiness
(see, for instance, Lombardi 2009).

\section{Derivation of the correction for galactic latitude}

The GALEX~AIS survey spans over a wide range of galactic latitudes (from $b_{gal} \simeq -5 ^{\rm o}$ to $b_{gal} \simeq -25 ^{\rm o}$) 
and it is expected that the stellar density decreases significantly over the surveyed area. As a result, in the evaluation of the
extinction:
\[
A_{\lambda}=\log(N^{*}_{\lambda}/N_{\lambda})/b_{\lambda}
\]
\noindent
$N^*_{\lambda}$ is a function of $b_{gal}$. In our work, we have used only a reference field, AIS189\_28, located at $b_{gal} = -6 ^{\rm o}$
hence it is necessary to determine the correction term, $\delta A_{lambda}$, such that,
\[
dA_{\lambda} (b_{gal}) = \frac {1}{b_{lambda}} \log \big( \frac {N^*(-6 ^{\rm o})}{N(b_{gal})} \big)
\]
\noindent
To derive it, we have used the infrared data and a simplified model of the galactic stellar density distribution.

LLA10 reported that the K-band star counts followed the galactic model. Both FM07 and Lombardi et al (2010)
designed procedures to correct for this gradient so they provide a reliable baseline to compare our correction with.
Note that no significant variations are expected in $b_{\lambda}$. Neither  the stellar density
varies significantly with galactic longitude in the anticenter region. 

The impact of this correction in extinction determinations is illustrated in Fig.~B1. Rough $A_K$ (without $\delta A_K$ correction)
have been computed by applying the star counts method on the sample constituted by {\bf all} {\it bona fidae} 2MASS sources in any given of the GALEX fields 
to the depth of the 2MASS survey (see Fig.~B1, top panel), as well as to a reduced sample including only the 2MASS sources with  NUV counterpart. Both panels display a clear gradient with galactic latitude, more pronounced in the top panel.
At the bottom, F07's infrared extinction map is shown for comparison (the map has been rebinned to the GALEX tiles size and distribution).

\begin{figure}
\includegraphics[width=6cm]{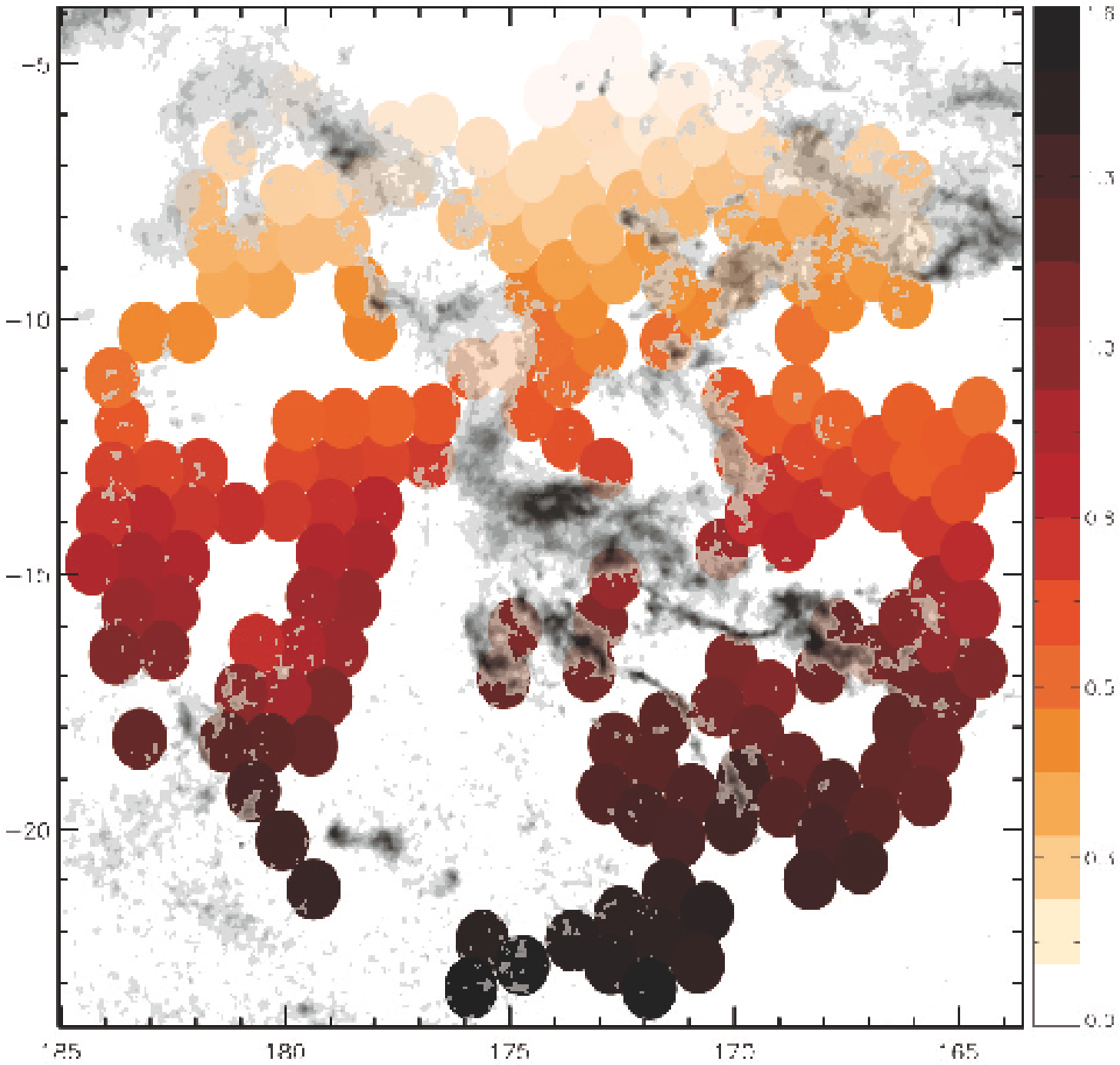} \\
\includegraphics[width=6cm]{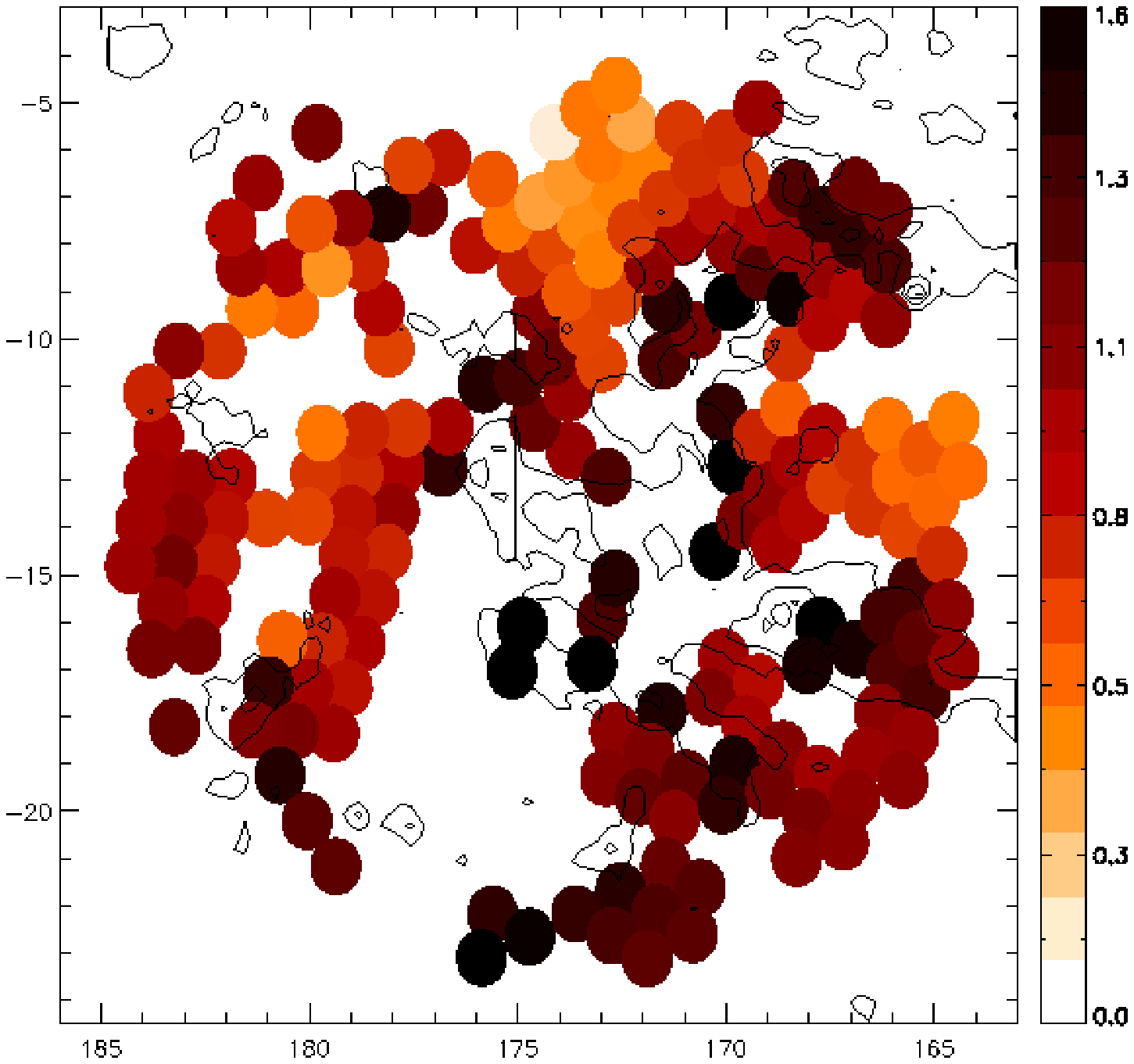}  \\
\includegraphics[width=6cm]{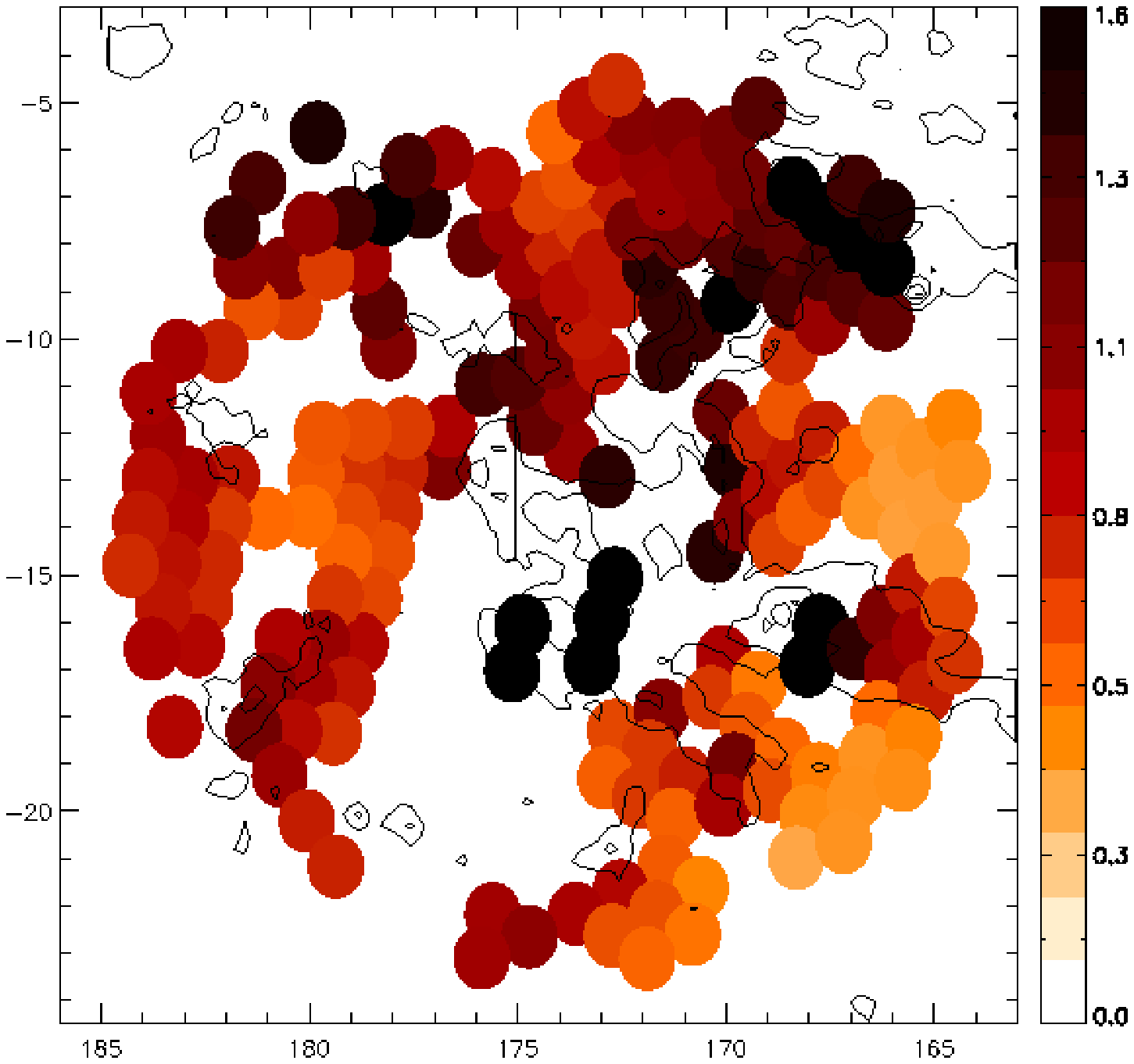} \\
\caption{K-band color-coded average extinction per GALEX~AIS tile in the TMC.
{\sl Top:} $\delta$A$_{\rm K}$ derived from all 2MASS sources per GALEX tile, 
{\sl Middle:} $\delta$A$_{\rm K}$ derived from stars with NUV counterpart, 
{\sl Bottom:} $\delta$A$_{\rm K}$ derived from binning F07 2MASS based extinction map. 
}
\label{extinction}
\end{figure}

According to the standard galactic model, the stellar density, $\rho ^*$, in the Galaxy can be parametrized in terms of a thin, 
$\rho^*_{a}$, and a thick disk, $\rho^*_{b}$, as follows:
\begin{equation}
\rho^* = \rho^*_{a}+\rho^*_{b}
\end{equation}
\noindent
with,
\begin{equation}
\rho^*_{a,b} = \rho^*_{0,(a,b)} \exp (-\frac{R}{L_{a,b}}) \exp (-\frac{z}{H_{a,b}})
\end{equation}
\noindent
with, $R$ and $z$ the galactocentric cylindric coordinates in the disk  (cylindric symmetry is assumed) and  $\rho^*_{0,a}$ and $\rho^*_{0,b}$ the volumetric density of stars in the center of the thin and thick disk, respectively. According to Ojha (2001), the parameters for the model are: $L_a = 2.8 \pm 0.3$~kpc, $L_b = 3.7 ^{+0.8}_{-0.5}$~kpc, $H_a = 250$~pc (Kuijken \& Gilmore 1989, Haywood 1997), $H_b = 860 \pm 200$~pc and $\rho^*_{0,b}/\rho^*_{0,a} = 0.035 \pm 0.02$. 
With these provisions, the total number of stars with absolute magnitude, $M$, observed within  a solid angle, $\Omega$, up to a given depth $r_d$, is given by, 
\begin{equation}
N(m) = \int _0 ^{r_d} \Phi (M) \rho^* \Omega r^2 dr
\end{equation}
\noindent
with $\Phi(M)$  the galactic distribution of stars with absolute magnitude, $M$. Let us assume that  $\Phi (M)$ can be consider constant and equal to $\Phi _0$ for all the sources in our survey. This would imply that all sources follow the simple  parametrization in Eq. (B2), irrespective of their nature, on scales of 1.4 deg (the resolution
of the survey).   With this approach, 
\begin{equation}
N(r_d) =  \Phi _0 \int _0 ^{r_d} \rho^* \Omega r^2 dr
\end{equation}
\noindent
Since the galactic longitude of our region of interest is $l_{gal} \simeq 180 ^{\rm o}$, $z \simeq (r+r_{\odot}) \tan (b_{gal})$, with $r_{\odot}$ the galactocentric distance of the Local Standard of
Rest (LSR)  and $(r,l_{gal},b_{gal})$ the LSR-centric galactic coordinates. Then,
\begin{equation}
\frac {dN(r_d)}{\cos b_{gal} db_{gal} dl_{gal}} = 
\end{equation}
\[
\Phi _0  (\exp (- \frac {r_{\odot}}{L_a}) \int _0 ^{r_d} \exp (-r(\frac {1} {L_a} + \frac {\tan b_{gal}}{H_a})) r^2 dr 
\]
\begin{equation}
+0.035 \exp (-\frac{r_{\odot}}{L_b} \int _0 ^{r_d} \exp (-r(\frac {1} {L_b} + \frac {\tan b_{gal}}{H_b})) r^2 dr ) 
\end{equation}
\noindent
Lets denote,
\begin{equation}
 S_{a,b} = \frac {1} {L_{a,b}} + \frac {\tan b_{gal}}{H_{a,b}}
\end{equation}
\noindent
Then,
\begin{equation}
\frac {dN(r_d)}{\cos b_{gal} db_{gal} dl_{gal}} = \Phi _0 \times
\end{equation}
\[
 (\exp (-r_{\odot}/L_a)\int _0 ^{r_d} \exp (-rS_a) r^2 dr
\]
\[ 
+ 0.035 \exp (-r_{\odot}/L_b) \int _0 ^{r_d} \exp (-rS_b) r^2 dr 
\]
\noindent
that depends on galactic latitude and depth of the survey. Let us denote,
\begin{equation}
 n(b_{gal},r_d) = \frac {dN(r_d)}{db_{gal} dl_{gal}}
\end{equation}
\noindent
and, 
\begin{equation}
dA_{\lambda} (b_{gal},r_d) = \frac {1}{b_{lambda}} \log \big( \frac {n^*(-6 ^{\rm o},r_d)}{n(b_{gal},r_d)} \big)
\end{equation}
\noindent
the effect of the stellar density gradient in the extinction derived from star counts.  Note that Eq.(B10) 
is parametrized in terms of the reference field AIS189\_28 (see Sect.~5), at $b_{gal} = -6 ^{\rm o}$.

In Fig.~B2, $dA_{\lambda} (b_{gal},r_d)$ is plotted for a range of galactic latitudes and possible depths
of the surveys according to Eq. (B10). To determine the depth of the survey (and the proper correction to be applied),
we have used three tiles, at different galactic latitudes:
the reference tile AIS189\_28 ($b_{gal} \simeq -6 ^{\rm o}$), AIS190\_42 ($b_{gal} \simeq -12 ^{\rm o}$) 
and AIS169\_28 ($b_{gal} \simeq -24 ^{\rm o}$). We have cross-checked that
$A_K = 0$ for all of them in the LLA10 extinction map.

We have used  $A_K$ measurements (see Fig.~B1) to evaluate 
$ dA_{\lambda} (-12 ^{\rm o},r_d) = A_{\lambda} (AIS190\_42) - A_{\lambda} (AIS189\_28)$ and
$ dA_{\lambda} (-24 ^{\rm o},r_d) = A_{\lambda} (AIS169\_28) - A_{\lambda} (AIS189\_28)$.
They are represented by dashed lines in Fig.~B2 as, $dA_{\lambda} (-24^{\rm o})$ and 
$dA_{\lambda} (-12 ^{\rm o})$. 

\begin{figure}
\includegraphics[width=8cm]{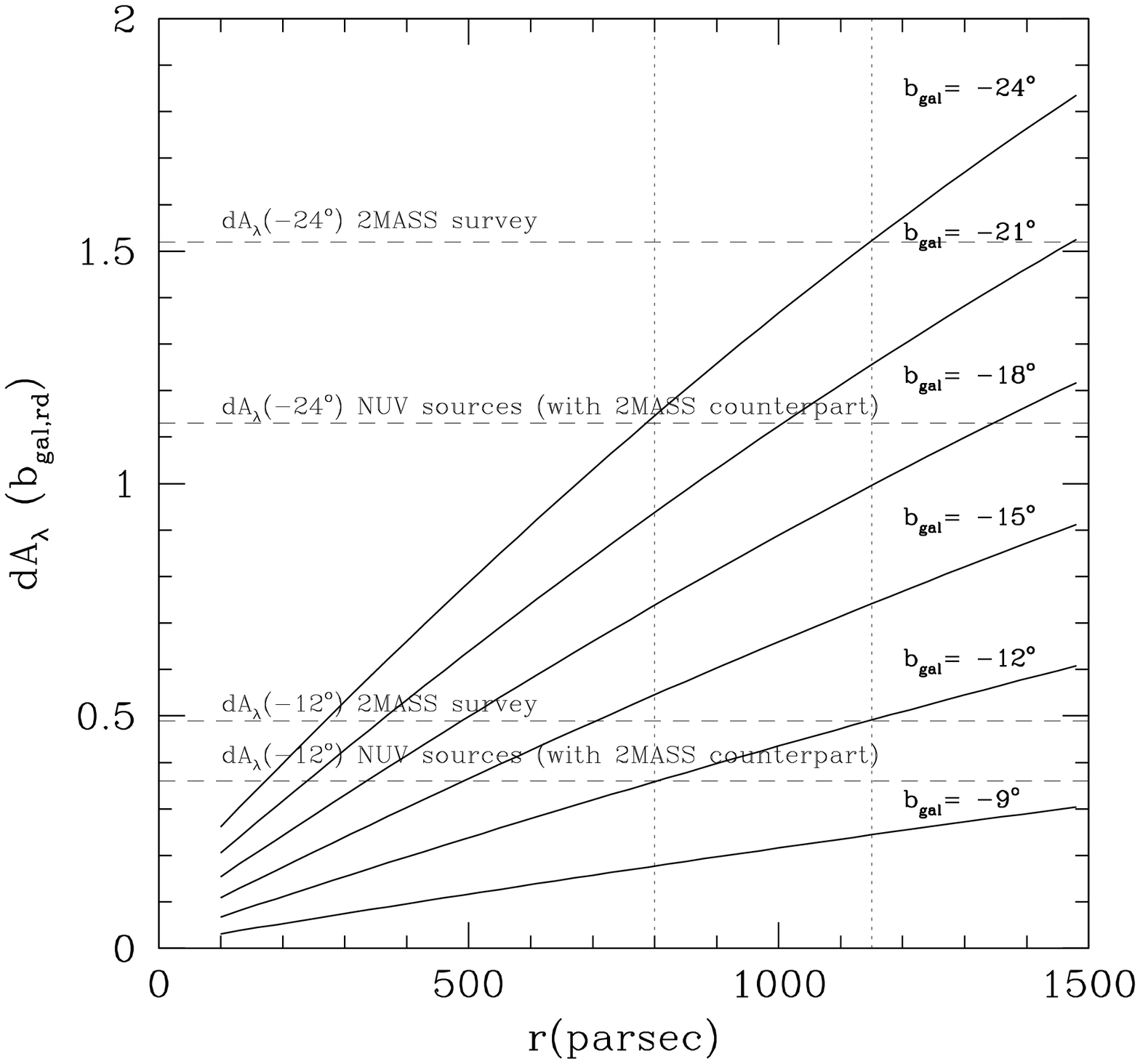}
\caption{Effect of the variation of stellar density with galactic latitude in the extinctions computed from star counts with
a given reference field at $b_{gal} = -6 ^{\rm o}$}
\label{galmod}
\end{figure}

As it is noticeable, the dashed lines meet the galactic model at
$r_d \simeq 800$~pc, for both galactic latitudes.  Thus, we assume that this is the depth of the
GALEX AIS survey in this area of the Galaxy.  From that, we compute the correction factor
to be applied to the measurements in Table~2 to be,
 
\begin{equation}
dA = (0.0640 \pm 0.0005) |b_{gal}(^o)| - (0.39 \pm 0.01)
\end{equation}
\noindent
(with $rms = 0.025$). The extinction map in the NUV band, after applying this correction, 
is displayed in Fig.~B3, for the record. The location of the obscuring clouds, 
the high latitude cirrus and the filaments are readily shown.
 
\begin{figure}
\includegraphics[width=8cm]{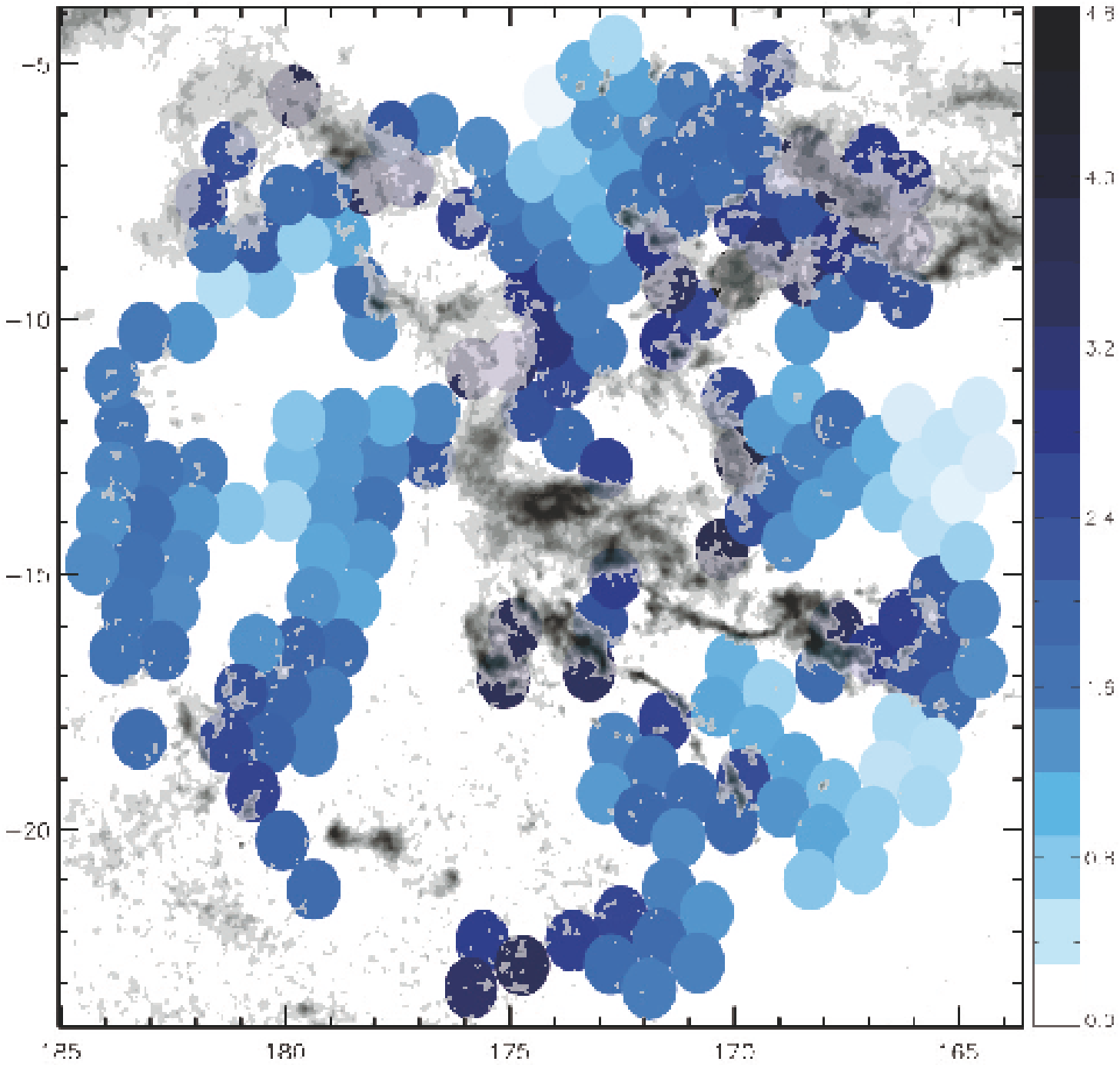} 
\includegraphics[width=8cm]{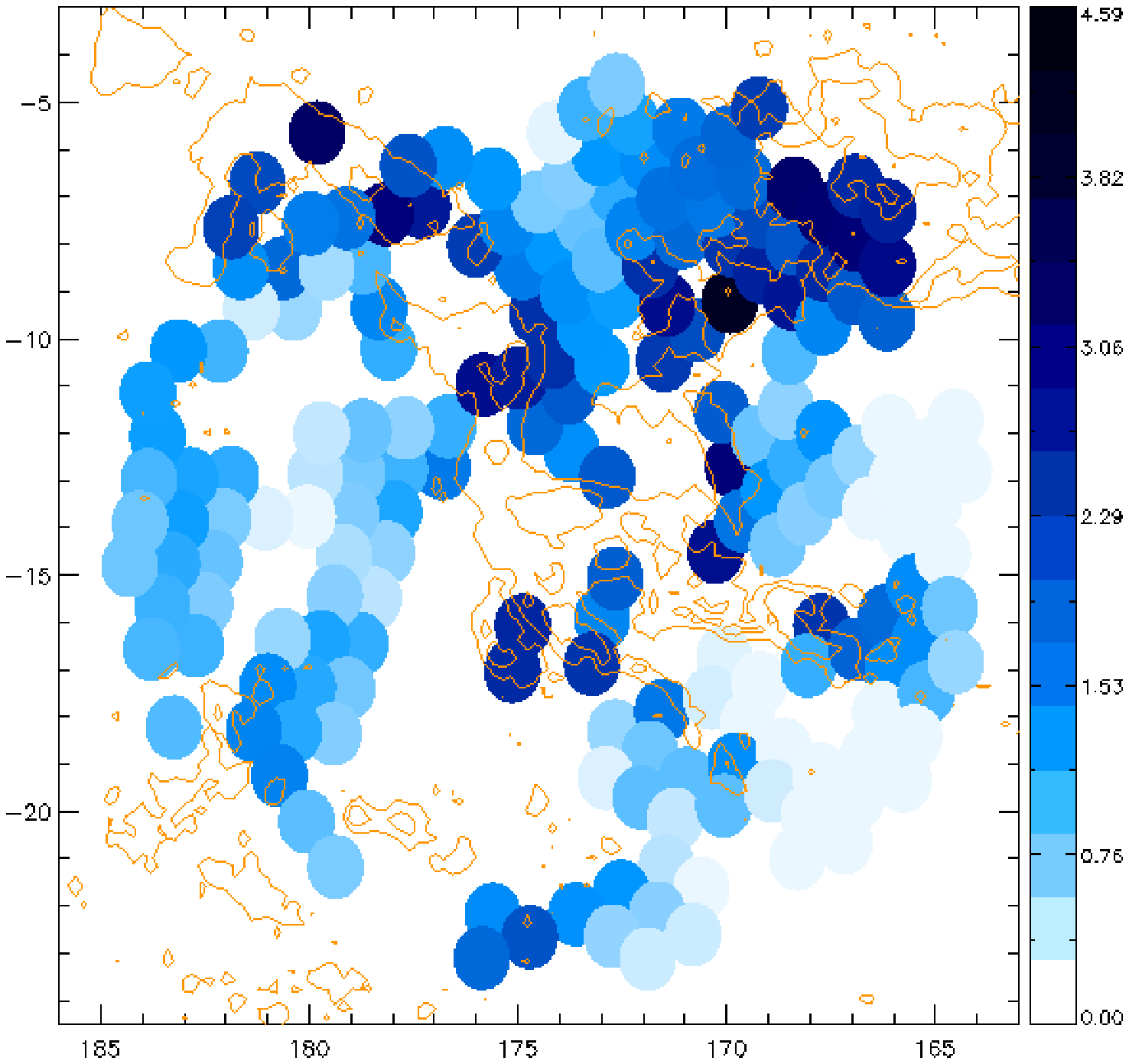}
\caption{Average extinction per GALEX tile in the NUV band determined from star counts. {\sl Top:} $A_{NUV}$ without applying the 
$\delta A_{NUV}$ correction, {\sl bottom:} $A_{NUV}$ after applying the correction.
The location of the obscuring clouds from F07  is outlined. Coordinates as in Fig~1.}
\label{galmod}
\end{figure}

\end{document}